\documentclass[floats,aps,prl,psfig,graphicx,twocolumn]{revtex}
\usepackage{graphicx}
\usepackage{amsmath}

\begin{document}
\title{Rocky core solubility in Jupiter and giant exoplanets}
\author{Hugh F. Wilson$^1$ and Burkhard Militzer$^{1,2}$} \affiliation{Departments of Earth
  and Planetary Science$^1$ and Astronomy$^2$, University of California Berkeley}

\author{Gas giants are believed to form by the accretion of hydrogen-helium gas around an 
initial protocore of rock and ice. The question of whether the rocky parts of the core dissolve into the fluid 
H-He layers following formation has significant implications for 
planetary structure and evolution. Here we use \emph{ab initio} calculations to study rock solubility in fluid hydrogen, choosing MgO as a representative
  example of planetary rocky materials, and find MgO to be highly soluble in H for temperatures in excess 
of approximately 10000~K, implying significant redistribution of rocky core material in Jupiter and larger exoplanets.}
\maketitle

The coming years will see a substantial increase in our understanding of giant planets, inside and outside our own solar system. The Juno mission to Jupiter will measure the gravitational field of our solar system's largest planet to unprecedented accuracy, while the Kepler mission and other planet-finding projects will greatly increase our statistical understanding of the mass/radius distribution of planets throughout the universe. Core-accretion models hold that giant planets form by the rapid runaway accretion of gaseous hydrogen-helium material from the protosolar nebula around a dense solid core once the core reaches a sufficiently large size. Understanding planetary formation therefore requires knowledge of the nature of giant planetary cores \cite{mizuno-78}. Measurements on Jupiter and Saturn can give us information about the internal mass distribution of these planets and hence provide information about their present-day core size, however it is poorly understood whether the initial protocore remains stable, or whether it partially or fully dissolves into the metallic hydrogen layers above and is redistributed throughout the envelope. Resolving the problem of core erosion and solubility may also be a factor in the observed enhancement of heavy elements in the outer layers of giant planet atmospheres which has been attributed to late-arriving planetesimals \cite{niemann-science-96,mahaffy}. 

Like other massive solid bodies in the outer solar system, the protocores of giant planets can be assumed to consist of a combination of rocky and icy materials. The rocky components are likely to be dominated by iron, magnesium, silicon and oxygen, and it was shown by Umemoto \emph{et al} \cite{umemoto-science-06} that MgSiO$_3$, a major constituent of the Earth's mantle, separates into SiO$_2$ and MgO at giant planet core conditions. Recent theoretical calculations \cite{wilson-erosion} predicted a substantial solubility of water ice in fluid hydrogen at the core/mantle boundary of Jupiter and Saturn, but the fate of  
less volatile rocky core components, however, remains unknown. The temperature and pressure conditions prevalent at giant planet core boundariesm on the order of 10 to 40 Mbar and 10000 to 20000~K, (higher for  super-Jupiters), are outside the range of laboratory experiments. As such, simulations based on \emph{ab initio} based theory remain the best available tools for studying solubility under these extreme conditions.

In this work we use free energy calculations based on density functional theory molecular 
dynamics (DFT-MD) and coupling constant integration (CCI) techniques to compute Gibbs free energies of pure and solvated hydrogen and MgO systems at giant planet core conditions. From these we determine the free energy of solvation of MgO in hydrogen and hence estimate the relationship between temperature and solubility of rocky core materials. As a necessary precursor, we investigate the ground-state crystal structure of MgO in the 10 to 40 Mbar regime where 
it has not previously been studied. Finally, we analyze the free energy of solubility in terms of its constituent thermodynamic quantities to determine the cause of the solubility behaviour determined in the calculations.

Gibbs free energies in this work were computed with a two-step CCI procedure
similar to that applied in several previous works \cite{wilson-erosion,wilson-prl-10,morales-pnas-09}. In CCI, the free energy of the system of interest, governed by a potential 
energy function $U_1 \left( \bf{r}_i \right)$, is connected via a 
thermodynamic integration to a simpler system with potential energy function 
$U_2 \left( \bf{r}_i \right)$. The Helmholtz free energy of the system of interest is then:

\begin{equation}
F_1 = \int_0^1 \langle U_1 - U_2 \rangle_\lambda d \lambda + F_2
\end{equation}

where the angle brackets denote an average taken over trajectories generated in the 
system governed by the hybrid potential energy function $U_\lambda = \lambda U_2 + \left( 1 - \lambda \right) U_1$. The Gibbs free energy is obtained from the Helmholtz free energy by the addition of $PV$.
The coupling constant integration in the present method is performed in two steps: the first from the system governed by density 
functional theory to a system governed by empirical potentials whose dynamics match, as closely as possible, 
those of the DFT system, and the second from the empirical potential system to an ideal system whose free 
energy is known analytically.

The classical potentials for the fluid systems were simple two-body potentials, fit to 
the interatomic forces of a DFT-MD run using the force-matching 
methodology of Izvekov \emph{et al} \cite{izvekov-jcp-04}. For the solid MgO system, we used a classical potential 
consisting of a combination of force-matched interatomic two-body potentials with a one-body harmonic term 
which anchors each atom to its ideal lattice position. Spring constants were fit to the mean square displacement of each atom from its lattice site during a trial DFT-MD run, then the two-body potentials generated from the DFT forces with 50\% of the harmonic terms subtracted. The second CCI step takes the system to an analytically known system, consisting of an ideal gas for the case of fluid systems, and a system of independent harmonic oscillators with spring constants as above for the solid systems.

The density functional theory calculations throughout this work were performed using the VASP code \cite{vasp}. We used pseudopotentials of the projector-augmented wave type \cite{paw}, the exchange-correlation functional of Perdew, Burke and Ernzerhof \cite{pbe}, a cutoff energy for the plane wave expansion of the wavefunctions of 900~eV, and a $2 \times 2 \times 2$ grid of k-points. The $\Delta G_{sol}$ values were confirmed to be well-converged with respect to these parameters to within the available error bars.

The phase diagram of MgO has been 
studied in detail up to pressures of approximately 5 Mbar by several authors \cite{belon-prb-10,tangney-jchemphys-09,zhang-grl-08} and it has been found to retain a CsCl crystal structure at these pressures, 
however the crystal structure at higher pressures has not to our knowledge been determined. Using the Ab Initio 
Random Structure Search method of Pickard and Needs \cite{pickard_needs} we searched for the most stable crystal structures of MgO at 
pressures of 10, 20, 30 and 40 Mbar, with one thousand possible structures explored at each pressure, and 
between one and four structural MgO units in each primitive cell.  For each pressure, we found the CsCl 
structure to be the lowest-enthalpy structure at 0K, with no other alternative structure being competitive in 
enthalpy. We also confirmed that MgO existed in the solid phase under all studied conditions by heating MgO to 40,000~K observing melting, equilibrating, and then re-cooling observing re-formation of the CsCl crystal structure. The Lindemann criterion \cite{lindemann-10}, defined as the 
ratio of the average mean square displacement of an atom from its lattice site divided by the nearest neighbour 
distance, was found to be 13.1\% for [40 Mbar, 20000K], 13.5\% for [20 Mbar, 15000K] and 13.4\% for [10 Mbar, 10000K], significantly smaller than the value of 18\% found to be required for the melting of MgO at pressures up to 3 Mbar by Cohen and Zong \cite{cohen-prb-94}, which further supports the existence of the material in the solid phase throughout the range of conditions studied.

Gibbs energies were computed for four systems: pure fluid hydrogen, fluid hydrogen containing
 one Mg atom, fluid hydrogen containing one O atom, and solid MgO in
the CsCl structure, at temperatures ranging from 5000 to 20000~K and
pressures ranging from 10 to 40 Mbar.  The stoichiometries used were H$_{128}$, H$_{127}$Mg, H$_{127}$O and MgO in a $3 \times 3 \times 3$ 54-atom unit cell. The Gibbs free energies obtained from the CCI simulations are shown in 
Table I. The error bars on the $G$ values are dominated by two terms, the more significant being the uncertainty in the volume at 
the desired pressure due to finite simulation time, and the other being the uncertainty in the $\langle 
U_{DFT} - U_{classical} \rangle$ terms in the thermodynamic integration. 

From these free energies, we obtain $\Delta G_{sol}(\text{MgO:254H})$ a free energy of solvation 
corresponding to the free energy change when one MgO unit is dissolved into a solution containing MgO at a 
concentration of one part in 254 atoms, 

\begin{equation}
\Delta G_{sol} (\text{MgO}:254\text{H}) = G(\text{H}_{254}\text{MgO}) - G(\text{H}_{254}) - G(\text{MgO}).
\end{equation}

Due to the large amount of hydrogen available in Jupiter and Saturn, a saturation solubility as low as one part in 254 is sufficient to allow the core to dissolve, provided that the core material can be redistributed away from the core-mantle boundary.
Neglecting the direct interaction between solute atoms (i.e. assuming the concentration is sufficiently low 
that direct solute-solute are sufficiently rare as to not have a significant effect upon the free energies) we 
may derive

\begin{equation}
G(H_{254}\text{MgO}) \approx G(\text{H}_{127}\text{Mg}) +G(\text{H}_{127}\text{O}) - 2kT \log(2),
\end{equation}

where the additional factor of $2kT \log(2)$ originates from the free energy of mixing.

Based on this formalism and the results in Table I, we obtain the free energies of solvation shown in Table \ref{table2}. 
Negative free energies indicate that solvation is preferred at a concentration of 1:254, while positive free 
energies indicate that the fluid system is supersaturated and that deposition of MgO, or 
formation of grains, will be thermodynamically favored. In contrast to the ice results of Ref. \cite{wilson-erosion} in which solubility became strongly favoured at relatively low temperatures of 2000-3000~K throughout the 10 -- 40 MBar range, we find MgO to become soluble at much higher T, but still well below Jupiter's core temperature, with the onset of significant solubilty lying in the eight to ten thousand Kelvin range. Solubility increases slightly with pressure, with a gradient of approximately 100~K per Mbar of pressure.

\begin{table}
\begin{tabular} {c |c c c c }
P(Mbar),& $G$(H$_{128}$) & $G$(H$_{127}$O) & $G$(H$_{127}$Mg) & $G$ (MgO) \\
T(1000K) & (eV) & (eV) & (eV) & (eV) \\
\hline
10, 10 & 214.9 $\pm$ 0.8 & 220.7 $\pm$ 1.1 & 241.1 $\pm$ 1.0 & 1024.2 $\pm$ 1.2 \\
20, 5 & 1173.6 $\pm$ 0.3 & 1195.9 $\pm$ 0.3 & 1223.0 $\pm$ 0.2 & 2300.4 $\pm$ 1.1 \\
20, 10 & 853.7 $\pm$ 0.3 & 872.0 $\pm$ 0.3 & 899.9 $\pm$ 0.3 & 2119.0 $\pm$ 0.6 \\
20, 15 & 474.9 $\pm$ 0.4 & 488.8 $\pm$ 0.5 & 516.9 $\pm$ 0.2 & 1884.1 $\pm$ 2.5 \\
30, 10 & 1343.8 $\pm$ 0.3 & 1372.6 $\pm$ 0.5 & 1405.7 $\pm$ 0.4 & 3011.0 $\pm$ 0.9 \\
40, 10 & 1755.9 $\pm$ 0.3 & 1793.9 $\pm$ 0.3 & 1831.5 $\pm$ 0.2 & 3775.1 $\pm$ 1.1 \\
40, 15 & 1403.6 $\pm$ 0.4 & 1437.7 $\pm$ 0.7 & 1475.7 $\pm$ 0.3 & 3566.4 $\pm$ 0.8 \\
40, 20 & 1014.3 $\pm$ 0.6 & 1043.9 $\pm$ 0.5 & 1082.1 $\pm$ 0.7 & 3335.6 $\pm$ 0.9 \\
\hline
\end{tabular}
\caption{Gibbs free energies of pure hydrogen, hydrogen with oxygen, hydrogen with magnesium, and solid MgO.}
\label{table1}
\end{table}

\begin{table}
\begin{tabular} {c c |r }
P [Mbar] & T [K]& $\Delta G_{sol}$ (eV) \\
\hline
10 & 10000 & -3.71 $\pm$ 2.2 \\
20 & 5000 & 4.35 $\pm$ 0.59 \\
20 & 10000 & -1.83 $\pm$ 0.75 \\
20 & 15000 & -8.26 $\pm$ 1.01 \\
30 & 10000 & -1.03 $\pm$ 0.9 \\
40 & 10000 & 0.02 $\pm$ 0.73 \\
40 & 15000 & -5.73 $\pm$ 1.04 \\
40 & 20000 & -12.72 $\pm$ 1.52 \\
\hline
\end{tabular}
\caption{ Gibbs free energies of solubility for MgO into hydrogen at a concentration of one part in 254 hydrogen atoms.}
\label{table2}
\end{table}

\begin{figure}[htbl]
\includegraphics[width=8.5cm]{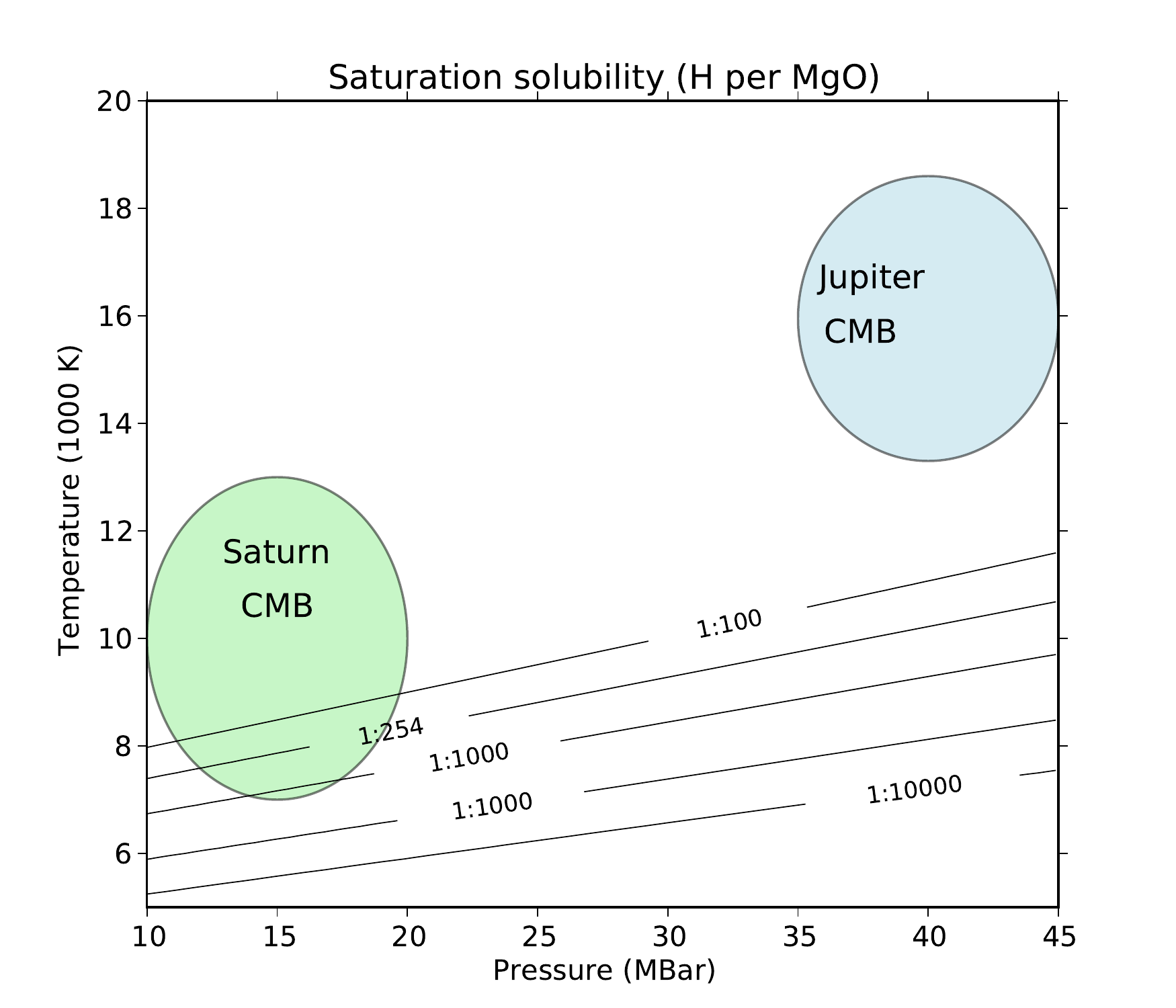}
\caption{Saturation solubility of MgO in H as a function of temperature and pressure. The estimated temperature/pressure conditions of the Jovian and Saturnian cores are 
shown for comparison. }
\label{figure1}
\end{figure}

These results may be generalized to solubility at other concentrations via consideration of the free energy of mixing \cite{wilson-erosion}. We explicitly neglect the interactions between solute atoms in the solution; this approximation will break down for high concentrations. Based a linear interpolation of the results in Table \ref{table2}, we estimated a saturation concentration for MgO in fluid hydrogen as a function of temperature and pressure throughout the 10-40 Mbar and 5000-20000K regime. A contour plot of constant saturation solubility is shown in Figure \ref{figure1}.  Saturation 
values above 1:100 are not shown, since the assumptions of non-interacting solute atoms break down rapidly as 
concentration increases.  While error bars are not shown, each contour should be considered to have an error bar of $\pm$ 1000 K. These results imply that MgO is highly soluble at temperature/pressure conditions corresponding to the Jovian core, but that estimates of Saturnian core conditions lie across a wide range of estimated concentrations from more than 1:10 to less than  1:1000. We thus estimate that MgO is likely to dissolve from Jupiter's core quite rapidly, while Saturn's core is significantly less soluble.  

As the Gibbs free energy is given by $G = U + PV - TS$, for fixed $P$ and $T$ the Gibbs free energy may be split into three components: an internal
energy component $\Delta U$, a volume component $P \Delta V$ and an
entropic component $T \Delta S$. The $P \Delta V$  values are easily extracted from the simulation, and we performed additional 8,000 step molecular dynamics simulations of each system to obtain accurate values of $\Delta U$. The 
$T \Delta S$ term is then the remaining term when the other terms are subtracted from $\Delta G$. The resulting breakdown as a function of temperature for the 20 and 40 Mbar 
results is shown in Figure 2. In all cases the $P \Delta V$ term shows a preference on the order of 2-4 eV for 
the undissolved MgO case, and the $\Delta U$ shows a somewhat larger preference in the same direction. The $T \Delta S$ term, however, favors solubility and shows a 
nearly linear dependence on temperature (implying an entropy gain associated with solvation of 
approximately 1.5 meV/K). In comparison with a similar breakdown of the solubility of water ice in  hydrogen in
Ref~\cite{wilson-erosion} we find that MgO has a stronger preference for the solid phase due not only to the stronger chemical bonding implied by the $U$ term but also due to $PV$ term; that is, the volume which Mg and O occupy is less when they are dissolved in hydrogen than when they are bonded into an MgO crystal. 

\begin{figure}[htbl]
\includegraphics[width=8.5cm]{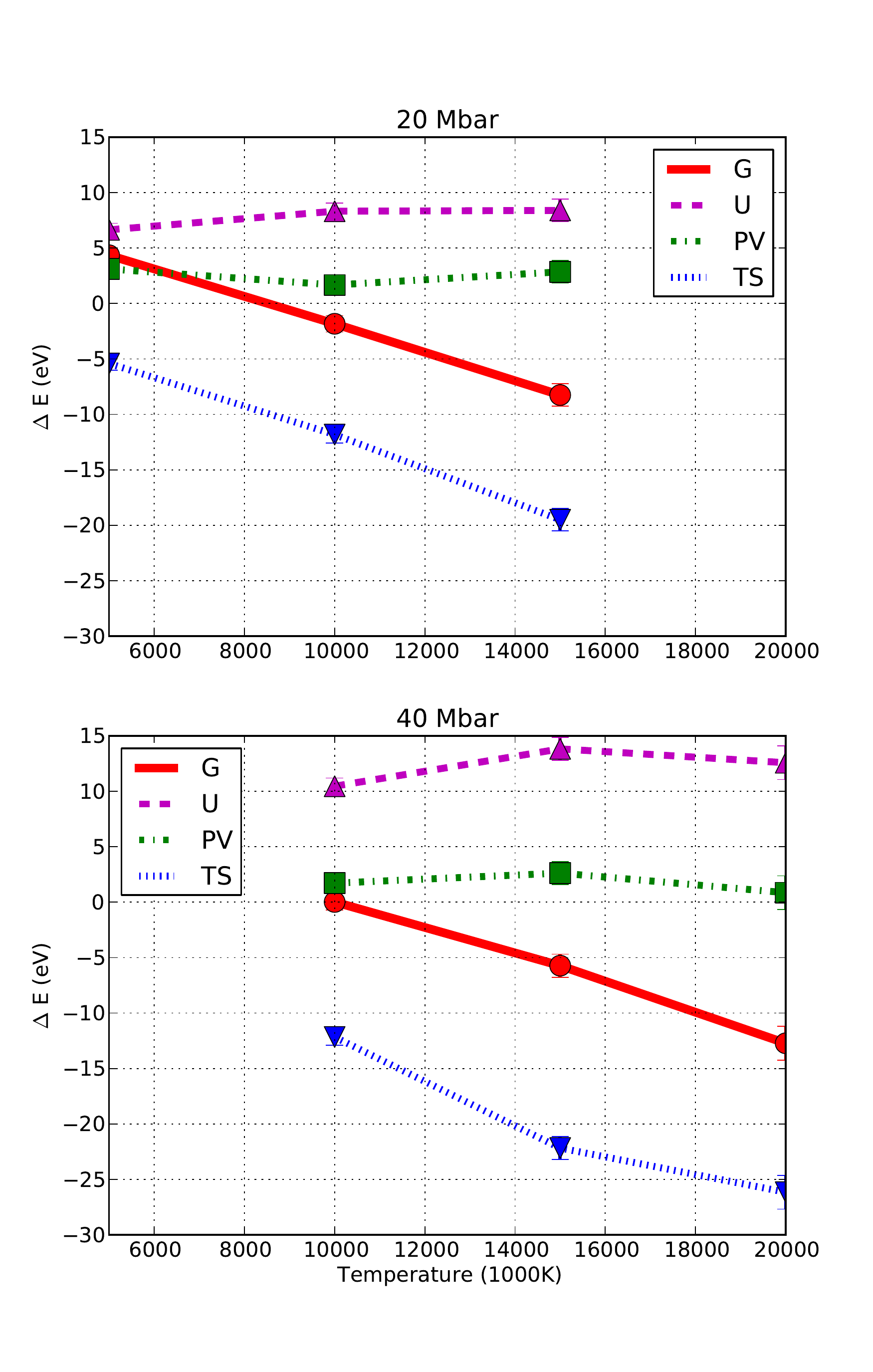}
\caption{(Color online) Breakdown of the computed $\Delta G$ values into its three thermodynamically 
constituent components: the $\Delta U$ term from the internal (chemical) energy, the $P \Delta V
$ term from pressure effects and the remainder $T \Delta S$ which represents entropic effects.}
\label{figure2}
\end{figure}

The results in this paper imply that MgO solubility at the core-mantle boundary will be significant in Jupiter and larger planets. Gas giant exoplanets larger than Jupiter will have hotter interiors, and hence can be expected to have even higher solubility.
Once core solubility becomes large, the rate at which core material can be redistributed throughout the bulk of the planet is limited by double diffusive convection \cite{guillot-book,stevenson-erosion}. Determination of the extent of this process is beyond the scope of this paper, however it was argued by Ref \cite{guillot-book} that the amount of core material redistributed over the lifetime of the planet could be on the orders of tens of percent for a planet of Jupiter size. We therefore assume that super-Jupiters are likely to have largely redistributed their initial protocores throughout the planet.

This initial study has simplified the problem in several ways. Firstly, we have assumed that Mg and O dissolve in a one-to-one ratio and ignored the possibility of a condensed phase with stoichiometry other than MgO being formed. While we consider this scenario unlikely on thermodynamic and chemical grounds, the possibility has not been excluded.  Furthermore, we have assumed dissolution of MgO into a pristine fluid layer containing hydrogen, rather than a fluid layer which may already have absorbed significant amounts of other dissolved core material. In particular, the oxygen concentration within the lower mantle may already be high due to solubility of ice and potentially SiO$_2$. While we expect the solubility of other rocky components to be broadly similar to that of MgO, detailed calculations on other core components such as SiO$_2$ may be valuable.

These results have substantial implications for models of Jupiter and large exoplanets. A non-uniform distribution of chemical elements in the interior, due to the limited rate at which
the core material can be redistributed, must be taken into account when modeling planetary interiors. Improved convective models which better predict the rate of core material redistribution within giant planets would be valuable, especially for the interpretation of the gravitational moments to be measured by the Juno probe. For large exoplanets exceeding Jupiter's mass, higher interior 
temperatures promote both solubility and redistribution, implying that the cores of sufficiently large super-Jupiters are likely to be completely redistributed. Additional advances in the 
spectroscopy of distant exoplanets raise the  possibility of relating composition to mass and hence 
detecting core erosion in distant exoplanets.

\bibliographystyle{h-physrev3} 

\bibliography{mgo}

{\bf Acknowledgements:}
This work was supported by NASA and NSF. Computational resources were supplied in part by TAC, NCCS and NERSC.

\end{document}